\documentclass[a4paper]{mn2e}
\usepackage{epsfig}

\def\msun{{\rm M_{\odot}}}
\def\rsun{{\rm R_{\odot}}}
\def\rl{{R_{\rm L}}}
\title[Black Widow Pulsars: the Price of Promiscuity]
{Black Widow Pulsars: the Price of Promiscuity}
\author[A.R.~King, M.B.~Davies and M.E.~Beer]{
A.~R.~King\thanks{E-mail: ark@astro.le.ac.uk}, M.~B.~Davies and M.~E.~Beer\\
Department of Physics and Astronomy, University of Leicester,
Leicester, LE1~7RH, UK\\
}

\volume{000}

\pagerange{\pageref{firstpage}--\pageref{lastpage}} \pubyear{2003}

\begin{document}

\maketitle

\label{firstpage}

\begin{abstract}
The incidence of evaporating `black widow' pulsars (BWPs) among all
millisecond pulsars (MSPs) is far higher in globular clusters than in the
field. This implies a special formation mechanism for them in
clusters. Cluster MSPs in wide binaries with WD companions
exchange them for turnoff--mass stars. These new companions eventually
overflow their Roche lobes because of encounters and tides. The
millisecond pulsars eject the overflowing gas from the binary, giving
mass loss on the binary evolution timescale. The systems are only
observable as BWPs at epochs where this evolution is slow, making the
mass loss transparent and the lifetime long. This explains why
observed BWPs have low--mass companions. We suggest that at least some
field BWPs were ejected from globular clusters or entered the field
population when the cluster itself was disrupted.

\end{abstract}


\section{Introduction}
It is widely believed that most millisecond pulsars (MSPs) have been
spun up by accretion from a close binary companion. This recycling
(Radhakrishnan \& Srinivasan, 1981) occurs when the neutron star
magnetic field has decayed to a relatively low value $\sim 10^8$~G. If
accretion ceases, the neutron star appears as a millisecond pulsar
with a very low spindown rate, as dipole radiation is very weak.

Not surprisingly, a large fraction of millisecond pulsars are observed
to be members of binary systems. In several cases the pulsars undergo
very wide eclipses, implying obscuration by an object considerably
larger than the companion star's Roche lobe. The obvious explanation
(Fruchter, Stinebring \& Taylor 1988) is that the obscuring object is
an intense 
wind from the companion star, driven in some way by energy injected by
the pulsar. In every case where eclipses are seen, both the
binary eccentricity and the pulsar mass function are extremely small,
giving companion masses $M_{2,{\rm min}} \la 0.1\,\msun$ in most cases (see Figs 1, 2).
Fig.~1 reveals another group of binary millisecond pulsars whose mass
functions are systematically lower, and it is natural to assume that
these are also evaporating systems with orbital inclinations which
prevent us seeing the eclipses (Freire et al., 2001).

\begin{figure}
  \begin{center}
    \epsfig{file=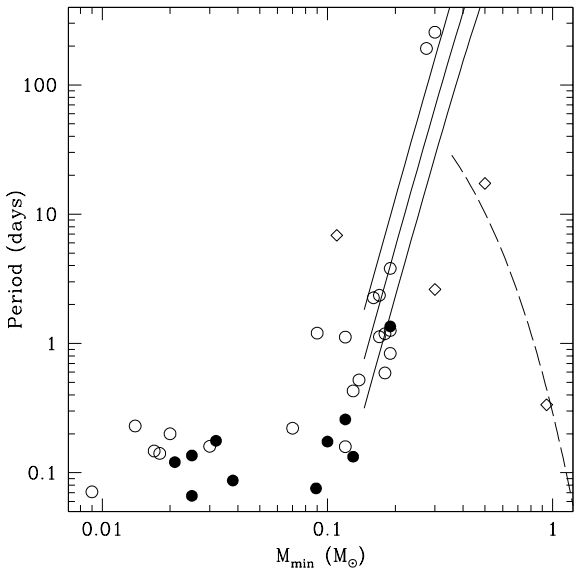, width=8cm}
  \end{center}
\caption{The observed period versus minimum secondary mass
distribution for binary MSPs in globular clusters. Diamonds represent
systems with eccentricities greater than 0.1. A filled symbol
indicates that the system has been observed to eclipse at some radio
frequency. The solid curves represent the theoretical distribution of
endpoints of late low-mass case B evolution for a range of
abundances. The upper and lower of the solid curves refer to
population I and population II stars respectively. The dashed curve
represents this distribution for early massive case B evolution. See
the Introduction for further details.}
\end{figure}

There have been numerous attempts to provide a coherent scheme for the
formation and evolution of these `black widow' pulsars (henceforth
BWPs). However these generally do not take account of the fact,
manifest from Figs 1 and 2, that the incidence of BWPs is considerably
higher among globular cluster MSPs than among field MSPs. (Rasio, Pfahl \&
Rappaport, 2000 consider the formation of short--period binary pulsars
in globulars, and note that most of them are BWPs, but do not consider
BWP formation explicitly.) Considering binary MSPs with $M_{2,{\rm min}} <
0.05\,\msun$, we find 11 BWPs among the 36 binary MSPs in globulars, but
only 2 BWPs out of 45 field binary MSPs. An MSP is almost 7 times more
likely to have a very low companion mass if it is in a globular
cluster rather than in the field.  Figures 1 and 2 show the observed
population of binary MSPs in globulars and the field respectively. The
globular population has been taken from the online catalogue of Paulo
Freire (http://www.naic.edu/$\sim$pfreire/GCpsr.html) and the field
from the review by Lorimer (2001). The figures also include the
theoretical period--mass relation for endpoints of late low--mass and
early massive case B evolution. The late low--mass case B endpoints
lie on the solid curves, and are taken from Rappaport et al. (1995)
\begin{equation}
P_{\rm orb} = 1.3 \times 10^5 M_{\rm wd}^{6.25} / (1 + 4 M_{\rm wd}^4)^{1.5} 
\,{\rm d} ~,
\end{equation}
where $P_{\rm orb}$ and $M_{\rm wd}$ are the period and white dwarf
mass (in solar units) 
respectively, and the upper and lower curves represent the spread
in the core mass--radius relation for both population I and population
II abundances respectively. Endpoints of
early massive case B evolution lie on the dashed curve, taken from
Taam, King \& Ritter (2000)
\begin{equation}
{\rm log} P_{\rm orb} \sim 2.56 - 3.1 M_{\rm wd} \,{\rm d} ~.
\end{equation}

These figures clearly show that there is some mechanism favouring the
production of BWPs in globular clusters. We suggest here that this is
the ability of binaries to exchange partners in globular clusters via
close encounters. This idea has far--reaching consequences.

It has been shown by Davies \& Hansen (1998) that 
exchange encounters in globular clusters will
tend to leave the most massive stars within binaries, independent of
the initial binary composition. When neutron stars exchange into
these binaries, the less massive of the two main--sequence stars
will virtually always be ejected. The remaining
main--sequence star will typically
have a mass of $\sim 1.5 -3 \ {\rm M}_\odot$. The binary will evolve into
contact once the donor star evolves up the giant branch.

\begin{figure}
  \begin{center}
    \epsfig{file=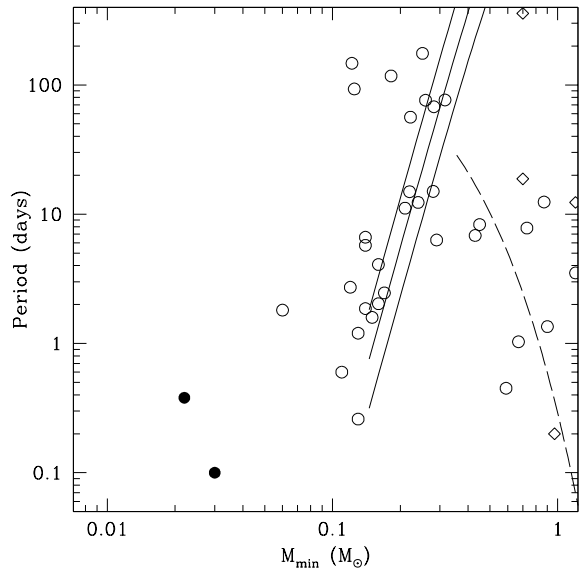, width=8cm}
  \end{center}
\caption{As for Figure 1 except for binary MSPs observed in
the field.}
\end{figure}

The subsequent evolution of such a system will depend on the mass
of the donor star and the separation of the two stars when the donor
fills its Roche lobe. For example, it has been suggested that the
system may enter a common envelope phase (eg Rasio et al. 2000).
Alternatively, the system may produce an {\sl intermediate--mass
X--ray binary (IMXB)}.
In such a system the neutron star may accrete
sufficient material (and with it, angular
momentum), for  it to acquire a rapid rotation (ie millisecond periods).
Because the donors
are all more massive than the present turn--off mass in globular
clusters, all IMXBs will have undergone their mass transfer
{\sl in the past}. If these systems evolve into MSPs, then we obtain,
quite naturally, what is observed today, namely a large MSP population
and a relatively small X--ray binary population.

Observations and modelling of the X--ray binary Cygnus X--2,
provide important clues in helping determine the subsequent evolution of
intermediate--mass systems.
This binary is unusual in that its donor has the appearance (by its
position in an HR diagram) of a slightly--evolved 3\,--\,5 M$_\odot$ star,
yet its measured mass is much lower ($\sim 0.5$ M$_\odot$).
The evolutionary history of Cyg X--2 has been considered
(King \& Ritter 1999, Podsiadlowski \& Rappaport 2000, and
Kolb et al. 2000).
The unusual evolutionary state of the secondary
today appears to indicate that the system has passed through a period
of high--mass transfer from an initially
 relatively--massive star ($\sim 3.6 $ M$_\odot$)
which had just evolved off the main sequence. The neutron star has somehow
managed to eject most of the $\sim 2 $ M$_\odot$ of gas transferred
at Super--Eddington rates from the donor during this phase. This evolutionary
history  may also apply to the IMXBs formed dynamically in globular clusters.
Vindication of this model also comes from studying the dynamical evolution
of the binary within the Galactic potential (Kolb et al. 2000).
A suitable progenitor binary originating in the
Galactic Disc has sufficient time, and could have received
a sufficient kick when the primary exploded to produce a neutron star,
to reach the current position of Cyg X--2.

\section{Widowhood}

Making millisecond pulsars requires two ingredients. First, the
neutron star must be spun up to millisecond periods, and second, it
must turn on as a radio pulsar. Adding a third ingredient -- a
companion which can be ablated -- makes it a black widow.

Recycling by accretion torques supplies the first ingredient. The
second is equivalent to demanding that this accretion should then
stop, to the point that the pulsar magnetosphere attains a state of
near--vacuum. No neutron star is observed both to accrete and pulse in
the radio: X--ray emission from radio pulsars is generally attributed
to mechanisms other than accretion (for the Be--pulsar binary PSR
1259-63 see e.g. King \& Cominsky, 1994, and Tavani \& Arons,
1997). Deep searches for pulsed radio emission from quiescent soft
X--ray transients containing neutron stars yield null results (Burgay
et al., 2003), despite the fact that the accretion rate, as measured
by the X--ray luminosity, drops as low as $\sim 5\times
10^{-15}~\msun~{\rm yr}^{-1}$. Evidently a spun--up neutron star can
appear as an MSP only if accretion effectively stops, and thus only if
the companion star detaches from its Roche lobe by many scaleheights.

The most obvious way of detaching is when the companion has a
core--envelope structure, and the latter is all lost. Thus in binaries
where a neutron star accretes from a low--mass giant (late low--mass
Case B evolution) we are eventually left with the low--mass helium
white--dwarf core of the giant in a wide detached binary whose
separation was set by the Roche lobe size of the giant just before it
lost its envelope. Many MSPs of this kind are seen, both in globulars
and the field (Figs. 1, 2). A second channel producing much tighter
and more massive MSP--WD binaries is early massive case B evolution
(King \& Ritter, 1999; Taam et al., 2000), where the evolved star was
initially more massive than the neutron star. Some of the field
distribution is consistent with this channel (Fig. 2). Van den Heuvel
\& van Paradijs (1988) proposed a third potential channel, to explain
an apparent lack of LMXBs below orbital periods of 3~hr. They
suggested that main--sequence companions might detach at this period,
as in cataclysmic binaries (see e.g. King, 1988 for a review of the
latter). However the most recent catalogue of Ritter \& Kolb (2003)
shows that neutron--star LMXBs have an apparently uniform distribution
at short periods, with no sign of a gap or lack of systems (cf King,
2003): of the 35 NLMXBs with known periods between 80 minutes and
10~hr, 8 lie between 80 minutes and 3~hr. It appears that the majority
of LMXBs do not detach at 3~hr, probably because their secondary stars
are somewhat nuclear--evolved (Schenker et al, in prep). Hence the
low--mass and intermediate--mass giant channels identified above
probably account for the production of most MSPs (Taam et al, 2000).

But this conclusion makes it difficult to supply the third BWP
ingredient listed above. If all binary MSPs emerge in binaries with
detached white dwarf companions, it is hard to see how efficient
ablation can occur and turn the system into a BWP. Observed MSP -- WD
binaries show no signs of incipient ablation. This is hardly
surprising given both the small target that the white dwarf presents
to the ablating radiation, and its high surface gravity, which must
reduce potential mass loss. At this point we recall the observation
above that the incidence of BWPs is far higher in globular clusters
than in the field.

\section{Companion exchange}

The obvious difference between field and cluster binaries is the
possibility of encounters and exchange interactions in clusters. A
cluster MSP in a binary with a white dwarf can exchange this
unpromising partner for one which is a potential BWP victim. The wide
low--mass MSP--WD remnants of low--mass Case B discussed above are
obvious candidates for this type of exchange. Their encounter
cross--sections are large, they are only loosely bound, and most
cluster stars are more massive than their white dwarfs and will thus
exchange into them in an encounter.

In globular clusters,
the timescale for a given binary to undergo an encounter
with a single star may be approximated as 

\begin{eqnarray}
\tau_{\rm encounter} & \simeq & 7 \times 10^{10} {\rm yr} \left( {10^5 \ {\rm pc}^{-3} \over
n } \right) \left( { V_\infty \over 10 \ {\rm km/s} } \right)
\ \nonumber \\ && \times
\left( { {\rm R}_\odot \over R_{\rm min} } \right) \left( { {\rm
M}_\odot \over M } \right) ~,\
\end{eqnarray}

\noindent 
where $n$ is the number density of single stars and $M$ is
the combined mass of the binary and a typical single star.  An
exchange encounter occurs when a third star, which is more massive
than the white dwarf, passes within a distance comparable to the
binary separation, ie $R_{\rm min} \simeq a$. For $a \sim 10$
R$_\odot$, and $n \sim 10^5$ stars pc$^{-3}$, $\tau_{\rm encounter}
\sim 2 \times 10^9$ yr. Hence the remnants of LMXBs are vulnerable to
exchange within the centres of virtually all globular clusters.  The
few, relatively long--period, systems shown in Fig. 1 are indeed
located either in the haloes of typical globular clusters or in the
cores of very low--density systems.

The product of an exchange encounter is an eccentric binary containing
the MSP and, typically, a main--sequence star close to the cluster
turn--off mass. The eccentricity is drawn from the distribution $f$
where $df/de = 2e$.  For wide binaries (periods around 100 days;
i.e. $a \sim 100 \,\rsun)$, subsequent encounters with single stars
perturb the binary, making it more bound, and leaving it with a new
eccentricity drawn from this distribution. Tidal circularization is
important for binaries with eccentricities close to unity, as the two
stars are much closer at periastron. If a binary circularizes, the
final separation, $a_{\rm f}=(1-e^2)a_{\rm ecc}$ where $a_{\rm ecc}$ is the
semi--major axis of the eccentric binary, having eccentricity $e$. The
timescale to achieve such circularization is a very sensitive function
of separation (e.g. Tassoul 1995). Observations reveal an absence of
eccentric binaries below periods of 10 days or so for stellar clusters
which are 1 Gyr old (Mathieu et al 1992). Given a sufficient number of
fly--by scatterings, we would therefore expect all wide binaries to be
left with very high eccentricities at some point, and for these
systems to then circularize with separations in the range $10 - 20
\,\rsun$.

Binaries which were initially much tighter (with periods around 10 days),
are likely to suffer the initial exchange encounter and perhaps only
one or two scatterings. In any case, these binaries are already tight
enough to circularize.

The outcome for virtually all binaries which have undergone exchanges
is thus a close binary (separation $a \sim 10 - 20 \,\rsun$) containing
an MSP in a circular orbit with a captured cluster star.  Subsequent
angular momentum loss via a magnetized wind is likely, driving the
system into contact (Verbunt \& Zwaan 1981). The initial mass $M_{2i}$
of the captured cluster star must be close to the turnoff value, as
the capture probability rises steeply with mass. This value is $\sim
0.7\,\msun$ for globular clusters in the present epoch, but was of
course higher in the past.

\section{Binary Bereavement}

In every case where an MSP is observed or inferred to ablate its
companion, the mass $M_2$ of the latter is much smaller than the
initial mass $M_{2i}$ inferred above. BWPs are apparently
observable only when ablation has reduced $M_2$ below $\sim
0.1\,\msun$. 

To investigate this we consider the evolution of the Roche lobe radius
$\rl \simeq 0.462(M_2/M)^{1/3}a$ of the companion, where $a$ is the
binary separation. We assume that the matter lost from this star and
thus from the binary carries specific angular momentum $\beta$ times
that of the companion $(M_1J/M_2M)$. Here $M_1 (\simeq 1.4\msun)$ is
the pulsar mass and $M = M_1 + M_2$. Evidently the value of $\beta$
depends on how the mass is lost. It is straightforward to show
(van Teeseling \& King, 1998) that
\begin{equation}
{\dot\rl\over\rl} = -{2\over t_J} + {\dot M_w\over
MM_2}\biggl[\biggr(2\beta - {5\over 3}\biggl)M_1 - M_2\biggr],
\end{equation}
where $t_J$ is the timescale for orbital angular momentum loss by
other means, e.g. gravitational radiation or magnetic braking. We
compare this change of $\rl$ with the change of the secondary's radius
$R_2$. We write $\dot R_2/R_2 = 1/t_{\rm exp}$ if the star expands
independently of mass loss, e.g. via thermal expansion across
the Hertzsprung gap or nuclear expansion, and
\begin{equation}
{\dot R_2\over R_2} = \zeta {\dot M_2\over M_2} 
\label{rdot}
\end{equation}
if the radius change results from mass loss. Here $\zeta$ is the {\it
effective} mass-radius index, i.e. the mass--radius index actually
followed by the companion along its evolutionary track. In general it
depends both on the nature of the secondary and the rate at which it
is losing mass, and must in general be calculated self-consistently
through the evolution. However we can infer its likely value quite
easily in many cases (see below). 

The evolution of the BWP binary depends on whether the mass loss
timescale $t_w = -M_2/\dot M_w$ is shorter or longer than $t_J$ or
$t_{\rm exp}$. Observations of the known BWPs imply that $t_w$ for
them is quite long ($\ga 10^9$~yr; Fruchter \& Goss, 1992; Stappers et
al., 2003), making their survival and observability
understandable. (Although orbital decay timescales of order $3\times
10^7$~yr are measured [Ryba \& Taylor, 1991] these must represent
short--term period derivatives of alternating sign rather than the
long--term evolutionary trend, a situation familiar for accreting
systems such as CVs and LMXBs.) Thus taking $t_w > {\rm min}(t_J,
t_{\rm exp})$ we see that the companion star comes into contact
with its Roche lobe on an angular momentum loss timescale $t_J$ or a
nuclear or thermal timescale $t_{\rm exp}$.

At this point, mass overflows the inner Lagrange point $L_1$. If
this mass were to accrete on to the neutron star, this would
extinguish the millisecond pulsar and turn the BWP into an LMXB. It
is instead more likely that the pulsar pressure is able to blow away
the matter flowing through $L_1$. We compare the pressure of the
pulsar wind at radius $R$
\begin{equation}
P_{\rm PSR} = {\dot E_{\rm rot}\over 4\pi R^2c}
\end{equation}
(where $E_{\rm rot}$ is the pulsar's rotational energy) with the ram
pressure of the matter trying to accrete at a rate $\dot M$
\begin{equation}
P_{\rm ram} = \rho v^2 < {\dot M\over 4\pi
R^2\Omega}\biggl({2GM_1\over R}\biggr)^{1/2}.
\end{equation}
Here $\rho, v$ are the density and infall velocity of matter at radius
$R$, and $\Omega$ is the solid angle subtended by this matter at the
pulsar. Thus
\begin{equation}
{P_{\rm PSR}\over P_{\rm ram}} > {\Omega\dot E_{\rm rot}\over c\dot
M}\biggl({R\over 2GM_1}\biggr)^{1/2}.
\end{equation}
Companions with the expected masses $\ga 0.7\msun$ fill the Roche lobe
at separations $a \sim 2\rsun$, so with $R$ of this order we find
\begin{equation}
{P_{\rm PSR}\over P_{\rm ram}} \ga 10{\Omega\dot E_{\rm rot, 35}\over \dot M_{-10}}
\end{equation}
where the units of $\dot E_{\rm rot}, \dot M$ are $10^{35}~{\rm
erg~s}^{-1}, 10^{-10}\msun {\rm yr}^{-1}$ respectively. Millisecond
pulsars have $\dot E_{\rm rot, 35} >1$, and we shall find $\dot
M_{-10} \sim 1$ below. Thus as the matter issuing through $L_1$
spreads out and tries to form a disc, $\Omega$ increases to the value
$\la 0.1$ where the pulsar pressure is able to blow it away. (In the
interval between submission and revision of this paper, the paper of
Sabbi et al., 2003 appeared, showing that just this appears to occur
in the globular cluster BWP~J1740-5340.)

This ejection increases $|\dot M_w|$, and thus affects the evolution
of the binary. If the mass--radius index $\zeta$ exceeds the critical
value
\begin{equation}
\zeta_{\rm crit} = \biggl(2\beta -
{5\over 3}\biggr){M_1\over M} - {M_2\over M} 
\label{stab} 
\end{equation}
the mass loss stabilizes when $\dot R_2 = \dot\rl$, so that
\begin{equation}
-\dot M_w = {2M_2\over (\zeta - \zeta_{\rm crit})t_J}
\label{aml}
\end{equation}
if evolution is driven by orbital angular momentum loss and
\begin{equation}
-\dot M_w = {M_2\over (\zeta - \zeta_{\rm crit})t_{\rm exp}}
\label{exp}
\end{equation}
if the evolution is instead driven by radius expansion independent of
mass loss (cf eqn. 22 of van Teeseling \& King, 1998). Note that,
unlike the case of mass {\it transfer}, a large companion mass tends
to stabilize mass loss, because it reduces the associated specific
angular momentum. Mass loss is stable even for a low--mass fully
convective or fully degenerate companion ($\zeta = -1/3$) provided
that $\zeta_{\rm crit} < -1/3$, i.e.
\begin{equation}
\beta < {2M - M_2\over 3M - 3M_2}.
\end{equation}
which approaches $\beta < 2/3$ for small $M_2/M$.  This process
remains stable even for small companion masses provided that $\beta
\la 2/3$. This value evidently requires that mass is lost from some
position inside the pulsar's Roche lobe. We shall investigate this
process in more detail in a future paper. We conclude quite generally
that an evaporating pulsar binary loses mass on a stellar expansion or
angular momentum loss timescale.

This conclusion poses stringent constraints on the observability of
such systems. The mass loss itself may obscure the pulsar and render
it impossible to detect, for example by increasing the dispersion
measure to the point that the pulses are smeared out. We note that in
several BWPs the pulsar is often obscured in this way for whole
orbital cycles (e.g. Lyne et al., 1990). This strongly suggests that
an evaporating pulsar binary is only detectable as a BWP at rather
small mass loss rates. We can quantify this effect by regarding the
outflow as a roughly spherical wind blown away from a region of the
size of the binary orbit. From the work of Burderi \& King (1994) the
free--free optical depth through this outflow at typical (400 -
1700~MHz) observing frequencies is $\ga 1$ for mass loss rates
\begin{equation}
\dot M_w > \dot M_{\rm crit} \simeq 10^{-11}T_6^{3/4}P_6\msun~{\rm
yr}^{-1}
\label{crit}
\end{equation}
(scaling their eqn 9 to general orbital periods $P_6 = (P/6~{\rm
hr})$, with the temperature in units of $10^6$~K, the expected value
for an ablated wind: Fruchter \& Goss, 1992). Note that Rasio, Shapiro
\& Teukolsky (1989) have to take a much cooler low--density wind in order to
reproduce the dispersion measure seen in PSR~1957+20. In the absence
of this constraint we can adopt a more reasonable wind temperature: cf
Wijers \& Paczy\'nski (1993).

This explains why BWPs are seen to have very small companion masses:
these evidently correspond to cases of rather low mass loss. Since the
mass loss timescale is also the evolution timescale, such systems also
have a long lifetime, increasing their discovery probability.
Evolution and mass loss are clearly slow once $M_2 \la 0.1\msun$ for
systems driven by angular momentum loss through gravitational
radiation. These account for all but one of the known BWPs. The other
BWP (J1740--53) has $P \simeq 1.35$~d, $M_{2,\rm min} \simeq
0.2\msun$ and is a subgiant (Ferraro et al., 2001). The
nuclear timescale of such a low mass subgiant is quite long, allowing
mass loss rates satisfying (\ref{crit}) here also; Watson et al (1985)
find a mass transfer rate $\sim 7.5 - 10 \times 10^{-11}\msun~{\rm
yr}^{-1}$ in the CV GK Per, which has almost the same period and secondary
mass, compared with $\dot M_{\rm crit} \simeq 6\times
10^{-11}T_6^{3/4}\msun~{\rm yr}^{-1}$ for these parameters.  In all
cases, detailed observational comparison evidently needs full binary
evolution calculations allowing for the change of $\zeta$ with
evolutionary state, as well as the full range of possible initial
companions. We shall present these in a later paper.

\section{BWPs in the field}

We have argued that in globular clusters, captures provide a mechanism
for making BWPs, whose formation is otherwise problematical. However
there are two BWPs in the field, namely 1957+20 and
2051--0827. Occam's razor suggests that we should not look for a
second formation mechanism, but a variation on the first. Two ideas
suggest themselves. Some BWPs made in globulars may be ejected by some
dynamical event (or the cluster itself was disrupted), 
or (rarely) captures may occur in the field. The latter is very unlikely
given the relatively low space density of stars in the field.

PSR~1957+20 has a tangential velocity of 190 kms$^{-1}$ (Arzoumian et
al., 1994) which corresponds to a space velocity $>$
220~km~s$^{-1}$. However its motion is restricted to the plane of the
Galaxy (its current galactic latitude is a mere $-5.2$
degrees). J2051--0827 is further out of the galactic plane ($b=-30.4$
degrees) but with a much smaller tangential velocity of 14 kms$^{-1}$
(Toscano et al 1999). It is unlikely, though not impossible, that both
objects formed from encounters in the field. It is much more probable
that these two systems have either been ejected from a cluster or that
the clusters themselves have been broken up. Metal--rich clusters
having kinematic properties similar to the thick disk are most likely
to be relevant here.

\section{Conclusions}

The formation of black widow pulsars has until now resisted simple
explanation. The problem is that formation requires two stages. A
binary companion is needed to spin up (recycle) the neutron star to
millisecond periods. But attempts to use the same companion as
the target which the pulsar ablates run into difficulty. A
semidetached binary always transfers some mass unless the donor
shrinks drastically. Thus neutron stars in short--period LMXBs are
always accreting, and therefore have difficulty in turning on as
millisecond pulsars and thus reaching the black widow stage.

Our answer to this difficulty uses the observed fact that the
incidence of BWPs in globulars is far higher than in the
field. Millisecond pulsars in `dead' wide binaries with WD companions
can exchange them for turnoff--mass stars. Encounters and tides bring
these new companions into tight orbits, where they eventually fill
their Roche lobes. The MSPs can eject from the binary gas overflowing
the inner Lagrange point, resulting in mass loss on the binary
evolution timescale. The systems are observable as BWPs only if this
timescale is long, making the mass loss transparent and the binary
lifetime long. This explains the preference of observed BWPs for
low--mass companions. We suggest that at least some field BWPs were
ejected from globular clusters, or were members of clusters that were
broken up.

A consequence of the picture suggested here is that a globular cluster
neutron star is likely to form a BWP rather than an LMXB if it
changes partners. Hence most LMXBs in globular clusters are still with
their original companions, and thus old. We shall investigate this
idea in a future paper.

\section{Acknowledgments} 

ARK and MBD are grateful to the organisers of the meeting on Globular
Clusters at the Kavli Institute for Theoretical Physics, Santa
Barbara, which provided the stimulus for this work. Theoretical
astrophysics research at Leicester is supported by a PPARC rolling
grant. ARK gratefully acknowledges a Royal Society Wolfson Research
Merit Award. MEB acknowledges the support of a UKAFF fellowship. We
thank the referee, Fred Rasio, for a very helpful report.

\label{lastpage}

\end{document}